\documentclass{appolb}

\begin{document}

\title{On the Finslerian extension of the Schwarzschild metric}

\author{Z.~K.~Silagadze
\address{Budker Institute of Nuclear Physics \and
Novosibirsk State University \\ 630 090, Novosibirsk, Russia }}
%\date{\today}

\maketitle

\begin{abstract}
We provide a Finslerian extension of the Schwarzschild metric based on 
heuristic arguments. The proposed metric asymptotically approaches not the 
Minkowski space-time but the Bogoslovsky locally anisotropic space-time
which arises naturally as a deformation of very special relativity.
\end{abstract}

%04.90.+e  - Other topics in general relativity and gravitation
%04.60.-m - Quantum gravity

\section*{}

Local Lorentz invariance is a fundamental cornerstone of modern phy\-sics.
Despite extremely tight experimental bounds on Lorentz symmetry violations
\cite{1} there still remains a tremendous interest in search for such
violations from both experimental and theoretical sides \cite{2}.

An ingenious way to break Lorentz invariance was suggested by Cohen and
Glashow \cite{3}. According to them, all current experimental limits on 
violations of Lorentz invariance may not imply that the full Lorentz group 
is the exact symmetry group of nature. The invariance with respect to its
four-parameter subgroup $SIM(2)$, which leaves invariant a preferred 
null-direction $n^\mu$, will suffice to meet all current experimental
limitations \cite{3,4}.  

Extended by space-time translations, $SIM(2)$ leads to the eight-parame\-ter
subgroup $ISIM(2)$ (first introduced, probably, in \cite{5}) of the 
Poincar\'{e} group which is supposed to be the genuine exact symmetry group 
of nature.

It is well known that, in isotropic situation,  there exists essentially only 
one way to generalize special relativity, namely by endowing space-time with 
some constant curvature \cite{6}. The resulting de Sitter and anti de Sitter
groups,  which are the most general isotropic relativity groups, can be 
considered as deformations of the Poincar\'{e} group \cite{7}. Similar 
deformations of $ISIM(2)$ were considered in \cite{8}. Among deformations of
$ISIM(2)$ particularly interesting is a one-parameter family of deformations,
called $DISIM_b(2)$, which does not leave invariant the standard Minkowski 
metric $ds^2=\eta_{\mu\nu}dx^\mu dx^\nu$ but the Finslerian metric
\begin{equation}
ds^2=(n_\sigma dx^\sigma)^{2b}(\eta_{\mu\nu} dx^\mu dx^\nu)^{1-b},
\label{eq1}
\end{equation} 
which was introduced by Bogoslovsky long ago \cite{9,10,11} in his attempts 
to formulate a generalization of the relativity theory for locally anisotropic
space-time. 

To investigate cosmological implications of the metric (\ref{eq1}), one 
should ask and answer the natural question: what is a curved-space 
generalization of this metric? The simplest heuristic guess applied in 
\cite{12P} is to change the Minkowski metric $\eta_{\mu\nu}$ in (\ref{eq1})
by the Robertson-Walker (or any other general relativistic) metric.

Although viable, such a procedure is somewhat ad hoc and there is no guarantee
that it gives the true solution. Of course, more systematic approach implies
Finslerian generalization of Einstein field equations. Some such 
generalizations were suggested (see, for example, \cite{12,13} and references 
therein) but at present it is not altogether clear \cite{14} whether any of 
them is on the correct route in the sense of representing reality.

In this note, applying another kind of heuristic arguments, we consider a
Finslerian generalization of Bogoslovsky type of the Schwarzschild metric
(units are such that $c=1$ and $G=1$)
\begin{equation}
ds^2=\alpha \; dT^2-\alpha^{-1}\;dR^2-R^2(d\theta^2+\sin^2{\theta}\,d\phi^2),
\label{eq2}
\end{equation}
where 
$$ \alpha=1-\frac{2m}{R}.$$
Earlier, a different type of Finslerian generalization of the Schwarzschild 
metric was considered by Asanov \cite{14P,15P}.

Firstly, let us concentrate on the radial part of the metric (\ref{eq2}).
In the Kruskal-Szekeres coordinates $t,x$ \cite{15,16} this radial part takes
the form
\begin{equation}
ds^2=\frac{32m^3}{R}\,e^{-R/2m}\,(dt^2-dx^2),
\label{eq3}
\end{equation}
where (in the exterior region $R>2m$)
\begin{equation}
\frac{t}{x}=\tanh{\left(\frac{T}{4m}\right)},
\label{eq4}
\end{equation}
and $R$ as a function of $x^2-t^2$ is implicitly determined through the 
relation
\begin{equation}
\left (\frac{R}{2m}-1\right)\,e^{R/2m}=x^2-t^2.
\label{eq5}
\end{equation}
Straightforward Finslerian generalization of the radial metric (\ref{eq3}) 
is
\begin{equation}
ds^2=\frac{32m^3}{R}\,e^{-R/2m}\,\left(\frac{dt-dx}{dt+dx}\right)^b\,
(dt^2-dx^2),
\label{eq6}
\end{equation}
where the the function $R(x,t)$ is implicitly determined through the 
relation
\begin{equation}
\left (\frac{R}{2m}-1\right)\,e^{R/2m}=\left(\frac{x-t}{x+t}\right)^b\,
(x^2-t^2),
\label{eq7}
\end{equation}
while the relation (\ref{eq4}) still remains valid.

Note that the r.h.s. of (\ref{eq7}) is invariant \cite{10} under the 
generalized Lorentz transformations (Bogoslovsky transformations) 
\begin{eqnarray}
x^\prime &=&e^{-b\psi}\,(x\,\cosh{\psi}-t\,\sinh{\psi}),\nonumber \\ 
t^\prime &=&e^{-b\psi}\,(t\,\cosh{\psi}-x\,\sinh{\psi}),
\label{eq8}
\end{eqnarray}
$\psi$ being the rapidity. Under the transformations (\ref{eq8}), the original
Schwarz\-schild coordinates $R$ and $T$ transform as follows
\begin{equation}
R^\prime=R,\;\;\;T^\prime=T-4m\psi.
\label{eq9}
\end{equation}
To rewrite the metric (\ref{eq6}) in the Schwarzschild coordinates we proceed
as follows. Let us introduce auxiliary variables $u=x+t$ and $v=x-t$. Then
(\ref{eq7}) takes the form
\begin{equation}
\alpha\,\frac{R}{2m}\,e^{R/2m}=u^{1-b}\,v^{1+b},
\label{eq10}
\end{equation}
which can be used to recast the metric (\ref{eq6}) as follows
\begin{equation}
ds^2=16m^2\alpha\,\left(\frac{du}{u}\right)^{1-b}
\left(-\frac{dv}{v}\right)^{1+b}.
\label{eq11}
\end{equation}
On the other hand, differentiating
$$\left(\frac{R}{2m}-1\right)\,e^{R/2m}=u^{1-b}\,v^{1+b}$$
and using (\ref{eq10}) in the result, we get
\begin{equation}
\alpha^{-1}\frac{dR}{2m}=(1-b)\,\frac{du}{u}+(1+b)\,\frac{dv}{v}.
\label{eq12}
\end{equation}
The second equation is obtained if we express (\ref{eq4}) in terms of $u$ and 
$v$ and differentiate the resulting equation. This gives
\begin{equation}
\frac{dT}{2m}=\frac{du}{u}-\frac{dv}{v}.
\label{eq13}
\end{equation}
Now we can solve (\ref{eq12}) and  (\ref{eq13}) to obtain
\begin{eqnarray} 
\frac{du}{u}&=&\frac{1}{2}\,\left[(1+b)\frac{dT}{2m}+\alpha^{-1}\frac{dR}{2m}
\right],\nonumber \\ 
-\frac{dv}{v}&=&\frac{1}{2}\,\left[(1-b)\frac{dT}{2m}-\alpha^{-1}\frac{dR}{2m}
\right].
\label{eq14}
\end{eqnarray}
Substituting these expressions into (\ref{eq11}), we finally get
$$ds^2=\alpha\,[(1+b)dT+\alpha^{-1}\,dR]^{1-b}\;
[(1-b)dT-\alpha^{-1}\,dR]^{1+b},$$
which can be rewritten as follows
\begin{equation}
ds^2=\left[\frac{(1-b)\,dT-\alpha^{-1}\,dR}{(1+b)\,dT+
\alpha^{-1}\,dR}\right]^b
[(1-b^2)\alpha\,dT^2-\alpha^{-1}\,dR^2-2b\,dT\,dR].
\label{eq15}
\end{equation}
Let us introduce a generalized tortoise coordinate
\begin{equation}
R_*=R+2m\,\ln{\left( \frac{R}{2m}-1\right)}+bT.
\label{eq16}
\end{equation}
Then
\begin{equation}
dR_*=\alpha^{-1}\,dR+b\,dT,
\label{eq17}
\end{equation}
and the metric (\ref{eq15}) takes a simpler form
\begin{equation}
ds^2=\left(\frac{dT-dR_*}{dT+dR_*}\right)^b\,\alpha\,(dT^2-dR_*^2),
\label{eq18}
\end{equation}
that is the same as
\begin{equation}
ds^2=[\sqrt{\alpha}\,(dT-dR_*)]^{2b}\,[\alpha\,(dT^2-dR_*^2)]^{1-b}.
\label{eq19}
\end{equation}
If the original Schwarzschild radial coordinate $R$ is restored through 
(\ref{eq17}), we get the final form of our radial Finsler metric 
\begin{equation}
ds^2=[(1-b)\,\alpha^{1/2}\,dT-\alpha^{-1/2}\,dR]^{2b}\,
[(1-b^2)\,\alpha\,dT^2-\alpha^{-1}\,dR^2-2b\,dR\,dT]^{1-b}.
\label{eq20}
\end{equation}
Of course, this can be obtained directly from (\ref{eq15}), but the detour 
through the tortoise coordinate makes it clear that it is $r=R+bT$, the 
asymptotic form of  this (generalized) tortoise coordinate, not $R$, which 
retains a well defined Finslerian asymptotic physical meaning. Indeed, 
$\alpha\to 1$ when $R\to\infty$, and in this limit (\ref{eq18}) takes just 
the Bogoslovsky form \cite{10}
$$ds^2=\left(\frac{dT-dr}{dT+dr}\right)^b\,(dT^2-dr^2).$$ 

But what about the angular part of the metric? Kruskal-Szekeres coordinates
are intimately related to the Fronsdal embedding of the Schwarzschild 
space-time into a six-dimensional pseudo-Euclidean space with signature
$(+,-,-,-,-,-)$ \cite{17}. The embedding is given by equations
\begin{eqnarray}
z_1&=&4m\sqrt{\alpha}\,\sinh{\left (\frac{T}{4m}\right )},\;\;
z_2=4m\sqrt{\alpha}\,\cosh{\left (\frac{T}{4m}\right )},\;\;
z_3=g(R),\nonumber \\
z_4&=&R\sin{\theta}\cos{\phi},\;\; z_5=R\sin{\theta}\sin{\phi},\;\;
z_6=R\cos{\theta},
\label{eq21}
\end{eqnarray}
where the function $g(R)$ satisfies \cite{17,18}
\begin{equation}
\left(\frac{dg}{dR}\right)^2=\frac{2m}{R}+\left(\frac{2m}{R}\right)^2+
\left(\frac{2m}{R}\right)^3=\alpha^{-1}\left[1-\left(\frac{2m}{R}\right)^4
\right]-1.
\label{eq22}
\end{equation}
Under this embedding, the Kruskal-Szekeres coordinates are given by \cite{19}
\begin{equation}
t=\frac{1}{4m}\sqrt{\frac{R}{2m}}\exp{\left(\frac{R}{4m}\right)}\,z_1,\;\;
x=\frac{1}{4m}\sqrt{\frac{R}{2m}}\exp{\left(\frac{R}{4m}\right)}\,z_2.
\label{eq23}
\end{equation}

For the needs of Finslerian generalization, we need the embedding to induce
the radial metric $(1-b^2)\,\alpha\,dT^2-\alpha^{-1}dR^2-2b\,dR\,dT$ instead
of Schwarzschildian $\alpha\,dT^2-\alpha^{-1}dR^2$. For this goal, we add
one more time-like dimension and modify the Fronsdal embedding in the 
following way
\begin{eqnarray}
z_0&=&bT,\;\;z_3=f(R),\;\;z_1=4m\sqrt{1-b^2}\,\sqrt{\alpha}\,
\sinh{\left (\frac{T}{4m}\right )}, \nonumber \\
z_2&=&4m\sqrt{1-b^2}\,\sqrt{\alpha}\,\cosh{\left (\frac{T}{4m}\right )},
\;\;z_4=(R+bT)\sin{\theta}\cos{\phi}, \nonumber \\ 
z_5&=&(R+bT)\sin{\theta}\sin{\phi},\;\;z_6=(R+bT)\cos{\theta},
\label{eq24}
\end{eqnarray}
where the function $f(R)$ is defined through
\begin{equation}
\left(\frac{df}{dR}\right)^2=\alpha^{-1}\left[1-(1-b^2)
\left(\frac{2m}{R}\right)^4\right]-1.
\label{eq25}
\end{equation}
Evidently, the embedding defined via (\ref{eq24}) and (\ref{eq25}) reduces 
to the Fronsdal embedding when $b=0$. The Finslerian variant of the  
Kruskal-Szekeres coordinates is given by
\begin{eqnarray}
t&=&\frac{1}{4m}\sqrt{\frac{R}{2m}}\exp{\left(\frac{R+bT}{4m}\right)}
\,\frac{z_1}{\sqrt{1-b^2}}, \nonumber \\
x&=&\frac{1}{4m}\sqrt{\frac{R}{2m}}\exp{\left(\frac{R+bT}{4m}\right)}
\,\frac{z_2}{\sqrt{1-b^2}},
\label{eq26}
\end{eqnarray}
so that (\ref{eq7}) is satisfied.

Let us now consider the Finslerian metric in the ambient space
\begin{equation}
ds^2=(N_A\,dz^A)^{2b}\,(\eta_{AB}\,dz^Adz^B)^{1-b},
\label{eq27}
\end{equation}
where the dummy indices run from $0$ to $6$ and 
$$\eta_{AB}=\mathrm{diag}(+1,+1,-1,-1,-1,-1, -1).$$ We want to choose seven 
unknown functions $N_A(R,T)$ in such a way that, in the case $\theta=\pi/2,\,
\phi=const$, the metric  (\ref{eq27}) to induce the radial metric (\ref{eq20})
under the embedding (\ref{eq24}). First of all we assume that the induced 
metric is axially symmetric with symmetry axis coinciding with the 
$z$-direction in the usual three-dimensional space, so that the metric does not
depend on the angle $\phi$. This demands $N_4=N_5=0$. The remaining five $N_A$
functions should ensure the above mentioned radial metric matching condition, 
along with the requirement that they determine the null-direction, 
$N_A\,N^A=0$. The resulting functional equations are
\begin{eqnarray}  
b(N_0-N_6)+\alpha^{1/2}\,\sqrt{1-b^2}\,\left(N_1\cosh{\frac{T}{4M}}-
N_2\sinh{\frac{T}{4M}}\right)=(1-b)\,\alpha^{1/2},&&\nonumber \\  
\left (\frac{2m}{R}\right)^2\sqrt{\frac{1-b^2}{\alpha}}\,\left (
N_1\sinh{\frac{T}{4M}}-N_2\cosh{\frac{T}{4M}}\right)-N_3\frac{df}{dR}-N_6=
-\alpha^{-1/2}, \hspace*{-2mm} &&  \nonumber \\  
N_0^2+N_1^2-N_2^2-N_3^2-N_6^2=0, \hspace*{30mm} && 
\label{eq28}
\end{eqnarray}
and they have the ``natural'' solution (as the case $b=1$ indicates)
\begin{eqnarray}
N_0&=&N_6,\;\;N_1=\sqrt{\frac{1-b}{1+b}}\,\cosh{\frac{T}{4m}},\;\;
N_2=\sqrt{\frac{1-b}{1+b}}\,\sinh{\frac{T}{4m}},\nonumber \\
N_3&=&\sqrt{\frac{1-b}{1+b}}, \;\;\;\;
N_6=\alpha^{-1/2}-N_3\,\frac{df}{dR}= \nonumber \\ & &
\alpha^{-1/2}\left[1-\sqrt{\frac{1-b}
{1+b}}\,\sqrt{\frac{2m}{R}}\,\sqrt{1-(1-b^2)\left(\frac{2m}{R}\right)^3}
\right].
\label{eq29}
\end{eqnarray}

We are almost done. Our generalization of the Schwarzschild metric (\ref{eq2})
is induced by (\ref{eq27}) on the Fronsdal-type sub-manifold parametrically
defined through (\ref{eq24}). At that, on this manifold,
$$\eta_{AB}\,dz^Adz^B=$$
\begin{equation}
(1-b^2)\,\alpha\,dT^2-\alpha^{-1}\,dR^2-2b\,dR\,dT-
(R+bT)^2\,(d\theta^2+\sin^2{\theta}\,d\phi^2),
\label{eq30}
\end{equation}
and
\begin{equation}
N_A\,dz^A=(1-b)\,\alpha\,^{1/2}\,dT-\alpha\,^{-1/2}\,dR+N_6\,d[(R+bT)
(1-\cos{\theta})].
\label{eq31}
\end{equation}
What remains is to show that this Finslerian generalization of the 
Schwarz\-schild metric is asymptotically the flat Bogoslovsky space-time. 
But in the limit $R\to\infty$, both $\alpha$ and $N_6$ tend
to unity, the generalized tortoise coordinate (\ref{eq16}) can be replaced
by $r=R+bT$ under differentials, and we have 
\begin{eqnarray} &&
\eta_{AB}\,dz^Adz^B \to dT^2-dr^2-r^2(d\theta^2+\sin^2{\theta}\,d\phi^2), 
\nonumber \\ &&
N_A\,dz^A \to dT-d(r\cos{\theta}).
\end{eqnarray}
Therefore, the asymptotic space-time has the Bogoslovsky metric (\ref{eq1}) 
with $n^\mu=(1,0,0,1)$.

The anisotropy parameter $b$ is expected to be very small and hard to access
experimentally \cite{8}. Why all the buzz then? The fact that $b$ arises
through the deformation of Cohen and Glashow's very special relativity 
indicates that the Bogoslovsky space-time parallels de Sitter (or anti de 
Sitter) space-time, not Minkowski space-time. Therefore $b$ parallels, in a 
sense, the cosmological constant and the question ``Why $b$ is so small'' is 
as fundamental as the cosmological constant problem \cite{8}. Maybe both 
questions are related to each other and have a common origin in quantum 
gravity.

\end{document}